\documentclass[aps,prl,twocolumn,superscriptaddress,nofootinbib,floatfix]{revtex4-2}
\usepackage{graphicx}
\usepackage{amsmath,amssymb}
\usepackage[colorlinks=true,citecolor=blue,linkcolor=blue,urlcolor=blue]{hyperref}

\begin{document}

\title{Spacetime torsion fixes the mass and spin of\\
gravitationally produced dark matter}

\author{Roh-Suan Tung}
\affiliation{Department of Physics and Center for Theoretical Physics, Chung Yuan Christian University, Taoyuan 320, Taiwan}
\affiliation{Institute of Advanced Studies, Nanyang Technological
University, 639673, Singapore}


\begin{abstract}
The gravitational production of dark matter from stochastic gravitational
waves requires the produced fermion to acquire a mass by unspecified
late-time physics. We show that this mass is supplied by spacetime torsion
alone---no Higgs sector and no free mass parameter. In the Quadratic Spinor
Lagrangian formulation of general relativity, extended to Einstein--Cartan,
a cosmological spinor condensate generates a vectorial trace torsion
$K\propto\dot\chi/\chi$; an explicit Clifford reduction confers on the
produced spin-$1/2$ fermion a pure Dirac mass
$M_{\rm eff}=(1/\sqrt6)\,|\dot\chi/\chi|$, with no pseudoscalar or cross
terms, locked to the Hubble rate at production,
$M_{\rm eff}\simeq(c_\chi/\sqrt6)H_*$. The relic abundance is then a
one-parameter prediction, $\Omega h^2\propto H_*^{5/2}$, and the spin is
fixed: the same framework admits no propagating spin-$3/2$ mode~\cite{nogo},
so the composite spin-$1/2$ Dirac fermion is its unique dark-matter
candidate.
\end{abstract}

\maketitle

Maleknejad and Kopp~\cite{MK} showed that a stochastic gravitational-wave
(GW) background breaks the conformal symmetry of massless Weyl fermions
and gravitationally produces them at one loop---a freeze-in origin for
dark matter (DM) from gravity alone. The mechanism leaves two inputs
unspecified: the GW source, and a means by which the produced fermions
acquire mass and become cold. Here we show that the Quadratic Spinor
Lagrangian (QSL) for general relativity~\cite{NT,TJ}, in its
Einstein--Cartan extension, supplies the second input geometrically: the
mass arises from spacetime torsion, with no Higgs sector and no free
mass parameter. Einstein--Cartan torsion has re-emerged as an active arena
for early-universe physics; it can, for instance, generate four-fermion
interactions that produce singlet-fermion dark matter over a wide mass
range~\cite{ECportal}. Here torsion plays a complementary role: it does not
\emph{produce} the dark fermion but supplies its mass and fixes its spin.

\paragraph{Setup.---}%
The QSL is the 4-form $\mathcal{L}=2\,D\Psi\,\gamma_5\,D\Psi$ ($\gamma_5$ the
chirality matrix), where $\Psi=\vartheta^a\gamma_a\psi$ is a
Clifford-algebra-valued 1-form built
from the coframe $\vartheta^a$ and a Dirac spinor $\psi$, and $D$ is the
covariant exterior derivative. By the spinor--curvature
identity~\cite{NTZ} this differs from the Einstein--Hilbert Lagrangian
only by a boundary term, so the QSL \emph{is} general relativity, with
the metric a spinor bilinear $g_{\mu\nu}=\bar\Psi_{(\mu}\Psi_{\nu)}$.
The object $\Psi_\mu=e^a_\mu\gamma_a\psi$ is a Dirac-vector; although such
a field carries spin-$3/2\oplus 1/2$ in general, the composite built from
a single spinor is \emph{purely} spin-$1/2$: its gamma-traceless part
vanishes identically,
\begin{equation}
\Psi^{(3/2)}_\mu=\gamma_\mu\psi-\tfrac14\gamma_\mu(\gamma^\nu\gamma_\nu)\psi=0 .
\label{eq:vanish}
\end{equation}
The propagating excitation is therefore a spin-$1/2$ Dirac field---the
same content as the Weyl fermion of~\cite{MK}.

Allowing torsion (the Einstein--Cartan extension), the spin-connection
field equation $\delta S/\delta\omega^{ab}=0$ fixes the torsion algebraically
in terms of the spinor, $T^a=f(\psi,D\psi,\vartheta)$. For a spatially
homogeneous condensate $\psi(t)=\chi(t)\hat\psi$ ($\hat\psi$ a constant
unit spinor) the covariant derivative
$D\psi=\chi\,(d\ln\chi+\omega)\hat\psi$ acquires a purely timelike piece
$d\ln\chi=(\dot\chi/\chi)\,dt$; isotropy then collapses the solution to a
vectorial trace torsion aligned with cosmic time,
\begin{equation}
K_a=K\,\delta^0_a,\qquad K\propto \dot\chi/\chi .
\label{eq:torsion}
\end{equation}

\paragraph{Geometric mass.---}%
Expanding the QSL about the Levi-Civita connection, the term quadratic in
the contorsion $\mathcal{K}=\tfrac14 K_{ab}\gamma^{ab}$ is algebraic in
$\Psi$ and has the structure of a mass term,
$\mathcal{L}_{K^2}=2\,\mathcal{K}\Psi\,\gamma_5\,\mathcal{K}\Psi$.
Writing $\mathcal{K}\Psi=\Xi$ as a spinor-valued 2-form and reducing the
Clifford contractions, $\mathcal{L}_{K^2}=\bar\psi\,\mathcal{S}\,\psi$
with
$\mathcal{S}=\tfrac12\epsilon^{pqmc}\overline{\Xi_{pq}}\gamma_5\Xi_{mc}$.
Projecting $\mathcal{S}$ onto the complete covariant basis
$\{\mathbf{1},\gamma_5,\gamma_a,\gamma_5\gamma_a,\gamma_{ab}\}$, the
pseudoscalar, vector, axial, and tensor channels vanish identically and
only the scalar survives:
\begin{equation}
\mathcal{S}=\tfrac23 K^aK_a\,\mathbf{1}
\;\Rightarrow\;
\mathcal{L}_{K^2}=\tfrac23 K^2\,\bar\psi\psi,\quad
M^2_{\rm eff}=\tfrac16\!\left(\frac{\dot\chi}{\chi}\right)^{\!2}.
\label{eq:mass}
\end{equation}
The mass-squared is read off in the vector-spinor normalization of the
QSL kinetic term, i.e.\ relative to $\bar\Psi_\mu\Psi^\mu=4\bar\psi\psi$;
the common factor $4$ cancels in the mass-to-kinetic ratio, leaving the
coefficient $\tfrac16$ exact. The explicit Clifford reduction
establishing $\alpha=\tfrac16$ and the vanishing of the pseudoscalar,
vector, axial, and tensor channels is given in the companion
paper~\cite{long} (appendix).
The result is a pure Dirac mass: the pseudoscalar ($\gamma_5$) channel
is absent and there are no cross terms. The clean vanishing of the other channels
is enforced by the cosmological symmetry: the source~\eqref{eq:torsion}
is the homogeneous, isotropic, timelike trace torsion, an $SO(3)$ scalar,
so the only invariant the bilinear can build is $K^aK_a$. A vector or
axial mass would require a preferred spatial direction, forbidden by
isotropy; the tensor channel would require the field's own spin current,
absent for a c-number background. For the same reason the result is a
scalar $\bar\psi\psi$ rather than the axial four-fermion contact term
familiar from Einstein--Cartan--Dirac theory, where the source is instead
the axial spin density of a propagating field.
This is the geometric ``drag'' mass: it vanishes in a static universe and
grows with the condensate's evolution rate.

\begin{figure}[t]
\centering
\includegraphics[width=0.92\columnwidth]{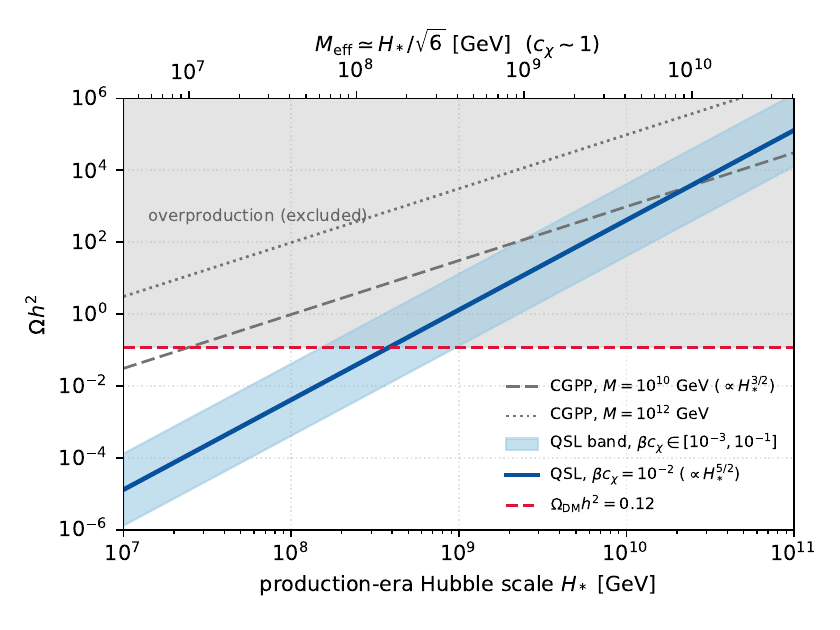}
\caption{Relic abundance vs.\ production scale $H_*$. Mass locking
collapses the two-parameter $(M,H_*)$ family of generic gravitational
production (grey, $\Omega h^2\propto H_*^{3/2}$) onto the single QSL
curve (blue, $\propto H_*^{5/2}$); the observed abundance (dashed) then
selects $H_*$ and the mass $M_{\rm eff}\simeq H_*/\sqrt6$ (top axis).}
\label{fig:relic}
\end{figure}

\paragraph{Mass locking and relic abundance.---}%
The condensate's evolution tracks the Hubble rate because no other scale
is available: when $\chi$ rolls on an asymptotically free
(Halpern--Huang) potential~\cite{HLT}, scale invariance is broken only by
the quantum trace anomaly, whose single emergent scale in an expanding
universe is $H$ itself. Hence $\dot\chi/\chi=c_\chi H$ with $c_\chi=O(1)$,
and the mass is locked to the expansion rate at the production epoch,
\begin{equation}
M_{\rm eff}=\frac{c_\chi}{\sqrt6}\,H_* ,
\label{eq:lock}
\end{equation}
the coefficient $1/\sqrt6$ being exact. Mass locking is the defining
prediction: the mass is not scanned independently of the cosmology. For
gravitational production with comoving density $n_*=\beta H_*^3$ ($\beta$
a dimensionless yield), the
relic abundance becomes a function of the single scale $H_*$,
\begin{equation}
\Omega h^2 \simeq 1.3\times10^{12}\,\beta c_\chi
\left(\frac{H_*}{10^{13}\,\mathrm{GeV}}\right)^{5/2},
\label{eq:relic}
\end{equation}
the exponent $\tfrac52$ (rather than the usual $\tfrac32$) being the
direct imprint of $M_{\rm eff}\propto H_*$. Matching
$\Omega_{\rm DM}h^2=0.12$ then fixes the production scale and the mass
(Fig.~\ref{fig:relic}); for $\beta c_\chi\sim10^{-2}$,
$H_*\sim10^{8}$--$10^{9}~\mathrm{GeV}$ and
$M_{\rm eff}\sim10^{8}~\mathrm{GeV}$. The number is illustrative; the
robust outputs are the locking relation~\eqref{eq:lock} and the
scaling~\eqref{eq:relic}. Mass locking further predicts a
gravitational-wave counterpart: $H_*$ fixes both $M_{\rm eff}$ and the
redshifted horizon-scale frequency $f_0\propto(H_*/M_{\rm Pl})^{1/2}$,
hence $f_0\propto\sqrt{M_{\rm eff}}$, with $f_0\sim0.2$--$0.7$~MHz over
the viable window---a stochastic signal in the ultra-high-frequency band of
bulk acoustic-wave and resonant-cavity detectors~\cite{UHF}, its peak
frequency locked to the dark-matter mass. The superfluid condensate's
quantum-turbulence relaxation adds a second, shape-distinct component
(a causal $f^3$ rise and Kolmogorov $f^{-8/3}$ fall); both lie below the
$\Delta N_{\rm eff}$ radiation bound, a target for the developing UHF
detector programme rather than current instruments.

\paragraph{Causality and outlook.---}%
The propagating spin-$1/2$ field obeys a second-order equation with the
metric d'Alembertian as principal symbol and so propagates causally on
the metric cone; the Velo--Zwanziger problem, a spin-$3/2$ pathology,
does not arise. If the Dirac-vector is instead promoted to an
\emph{independent} spin-$3/2$ field, no propagating massive mode results:
the QSL action depends on the field only through the metric
$g=\Psi\otimes_S\Psi$ and the scalar $\bar\psi\psi$, so its quadratic
fluctuations factor through a massless graviton and scalar, with every
pole on the metric cone~\cite{nogo}. The composite spin-$1/2$ fermion is
thus the unique propagating candidate, and the spin-$3/2$
pathologies---Velo--Zwanziger and catastrophic
production~\cite{raritron}---are precluded along with the field that would
suffer them. Through the QSL's super-$SL(2,\mathbb{C})$ structure this
surviving mode is the Goldstino of the local supersymmetry broken by the
metric condensate, placing the candidate on the oldest branch of
supersymmetric dark matter---the light gravitino of Pagels and
Primack~\cite{PP}---but as a composite, gravitational realization carrying a
geometric mass. Embedded in an asymptotically free scalar-field
cosmology~\cite{HLT}, the scale breaking that sets $\dot\chi/\chi$ is the
trace anomaly, geometrically carried by the trace torsion. The companion
papers develop these results in detail~\cite{long,nogo}.

\paragraph{Conclusion.---}%
The Quadratic Spinor Lagrangian supplies, from spacetime torsion, the
mass that gravitational dark-matter production otherwise leaves
unexplained. The mass is geometrically fixed and locked to the Hubble
scale, turning the relic abundance into a one-parameter prediction and
eliminating the need for a separate Higgs sector.

\begin{acknowledgments}
The author thanks J.~M.~Nester for discussions on the quadratic spinor
Lagrangian, and acknowledges the late Kerson Huang for the collaboration
on asymptotically free scalar-field cosmology.
\end{acknowledgments}


\begin{thebibliography}{9}
\bibitem{MK} A.~Maleknejad and J.~Kopp,
Phys. Rev. Lett. \textbf{136}, 131501 (2026).
\bibitem{NT} J.~M.~Nester and R.-S.~Tung,
Gen. Relativ. Gravit. \textbf{27}, 115 (1995).
\bibitem{TJ} R.-S.~Tung and T.~Jacobson,
Class. Quantum Grav. \textbf{12}, L51 (1995).
\bibitem{NTZ} J.~M.~Nester, R.-S.~Tung, and V.~V.~Zhytnikov,
Class. Quantum Grav. \textbf{11}, 983 (1994).
\bibitem{ECportal} M.~Shaposhnikov, A.~Shkerin, I.~Timiryasov, and
S.~Zell, Phys. Rev. Lett. \textbf{126}, 161301 (2021);
\textbf{127}, 169901(E) (2021).
\bibitem{raritron} K.~Kaneta, W.~Ke, Y.~Mambrini, K.~A.~Olive, and
S.~Verner, Phys. Rev. D \textbf{108}, 115027 (2023).
\bibitem{HLT} K.~Huang, H.-B.~Low, and R.-S.~Tung,
Class. Quantum Grav. \textbf{29}, 155014 (2012).
\bibitem{PP} H.~Pagels and J.~R.~Primack,
Phys. Rev. Lett. \textbf{48}, 223 (1982).
\bibitem{UHF} N.~Aggarwal \textit{et al.},
Living Rev. Relativ. \textbf{24}, 4 (2021);
``Challenges and Opportunities of Gravitational-Wave Searches above 10 kHz,''
arXiv:2501.11723 (2025).
\bibitem{long} R.-S.~Tung, ``Dark matter from the quadratic spinor
Lagrangian. I. Geometric mass for a gravitationally produced spin-1/2
fermion,'' (2026).
\bibitem{nogo} R.-S.~Tung, ``Dark matter from the quadratic spinor
Lagrangian. II. A spin-3/2 no-go and the uniqueness of the spin-1/2
candidate,'' (2026).
\end{thebibliography}
\end{document}